\documentclass[aps,prb,twocolumn,superscriptaddress,showpacs]{revtex4-2}
\usepackage{amsmath,amssymb,wasysym,graphicx}
\pdfoutput=1
\usepackage[utf8]{inputenc}
\usepackage[T1]{fontenc}
\usepackage{xcolor}

\IfFileExists{newtxtext.sty}
   {\usepackage{newtxtext,newtxmath}}
   {\IfFileExists{stix.sty}
      {\usepackage{stix}}
      {\IfFileExists{mathptmx.sty}
      {\usepackage{mathptmx}}{} } }

\usepackage{textcomp}

\usepackage{bm}

\IfFileExists{siunitx.sty}{\usepackage{booktabs,siunitx}}{}

\usepackage{color}
\definecolor{LinkColor}{rgb}{0.256,0.439,0.588}
\usepackage{hyperref}
\hypersetup{
   colorlinks=true,
   citecolor=LinkColor,
   linkcolor=LinkColor,
   urlcolor=LinkColor
}

\usepackage{pifont}
\usepackage[normalem]{ulem}

\usepackage{multirow}

\begin{document}

\title{Caution on Gross-Neveu criticality with a single Dirac cone: \\Violation of locality and its consequence of unexpected finite-temperature transition}

\author{Yuan Da Liao}
\affiliation{State Key Laboratory of Surface Physics, Fudan University, Shanghai 200438, China}
\affiliation{Center for Field Theory and Particle Physics, Department of Physics, Fudan University, Shanghai 200433, China}

\author{Xiao Yan Xu}
\affiliation{Key Laboratory of Artificial Structures and Quantum Control (Ministry of Education), School of Physics and Astronomy, Shanghai Jiao Tong University, Shanghai 200240, China}

\author{Zi Yang Meng}
\email{zymeng@hku.hk}
\affiliation{Department of Physics and HKU-UCAS Joint Institute of Theoretical and Computational Physics, The University of Hong Kong, Pokfulam Road, Hong Kong SAR, China}

\author{Yang Qi}
\email{qiyang@fudan.edu.cn}
\affiliation{State Key Laboratory of Surface Physics, Fudan University, Shanghai 200438, China}
\affiliation{Center for Field Theory and Particle Physics, Department of Physics, Fudan University, Shanghai 200433, China}
\affiliation{Collaborative Innovation Center of Advanced Microstructures, Nanjing 210093, China}

\begin{abstract}
Lately there are many SLAC fermion investigations on the (2+1)D Gross-Neveu criticality of a single Dirac cone. While the SLAC fermion construction indeed gives rise to the linear energy-momentum relation for all lattice momenta at the non-interacting limit, the long-range hopping and its consequent violation of locality on the Gross-Neveu quantum critical point (GN-QCP) -- which a priori requires short-range interaction -- has not been verified. Here we show, by means of large-scale quantum Monte Carlo simulations, that the interaction-driven antiferromagnetic insulator in this case is fundamentally different from that on a purely local $\pi$-flux Hubbard model on the square lattice. In particular, the antiferromagnetic long-range order has a finite temperature continuous phase transition, which appears to violate the Mermin-Wagner theorem, and smoothly connects to the previously determined GN-QCP. The magnetic excitations inside the antiferromagnetic insulator are gapped without Goldstone mode, even though the state spontaneously breaks continuous $SU(2)$ symmetry. These unusual results point out the fundamental difference between the QCP in SLAC fermion and that of GN-QCP with short-range interaction. 
\end{abstract}

\date{\today}

\maketitle

\section{Introduction}
Massless Dirac fermions are ubiquitously present as the low-energy description of many condensed matter systems including graphene~\cite{castroElectronic2009}, twisted bilayer graphene~\cite{xuKekule2018,liaoValence2019,liaoCorrelated2021,liaoCorrelation2021}, d-wave superconductors~\cite{wenTheroy1996,kimMassless1997,leeU12005,leeDoping2006}, algebraic spin liquid~\cite{hermeleAlgebraic2005,kimMassless1997,ranProjected2007,hermeleErratum2007,xuMonte2019,wenTheroy1996,dupuisAnomalous2021,calveraTheory2021,liaoGross2022,assaadPhase2005} and the deconfined quantum criticality~\cite{maDynamical2018,qinDuality2017,senthilQuantum2004,liaoDirac2022,liaoPhase2022,satoDirac2017,liuMetallic2022,liuSuperconductivity2019,wangPhases2021,wangDoping2021,liDeconfined2019a,zhuQuantum2022,liuDoes2023}; in high-energy physics, the dynamical
massless Dirac fermions in quantum chromodynamics and the existence of a deconfined phase in compact quantum electrodynamics have attracted great attentions and remains unsolved~\cite{fiebigMonopoles1990,herbutPermanent2003,hermeleAlgebraic2005,nogueiraCompact2008,karthikNumerical2019,karthikQED32020,calveraTheory2021,albayrakBootstrapping2022}. 
Nonetheless, it is generally believed that strong local interactions can generate a finite mass for the Dirac
fermions and spontaneously result in a quantum phase transition~\cite{sorellaSemi1992,mengQuantum2010,changQuantum2012,otsukaUniversal2016,toldinFermionic2015,liuDesigner2020}. The corresponding quantum critical point (QCP) are typically described by the Gross-Neveu (GN)
university classes~\cite{grossDynamical1974,boyackQuantum2021}. 
In particular, a single Dirac cone, realized in the the SLAC fermion model with long-range hopping in (2+1)D~\cite{drellStrong1976,karstenVacuum1979}, was found to give rise to an Ising-type ferromagnetic order that generates a $Z_2$ symmetry-breaking mass
gap~\cite{tabatabaeiChiral2022}, or an antiferromagnetic Mott insulator that breaks the $SU(2)$ spin rotational symmetry~\cite{langQuantum2019}. The associated QCPs from Dirac semimetal (DSM) to insulators are believed to belong to the (2+1)D chiral Ising or Heisenberg GN universality classes.

\begin{figure}[htp!]
\includegraphics[width=\columnwidth]{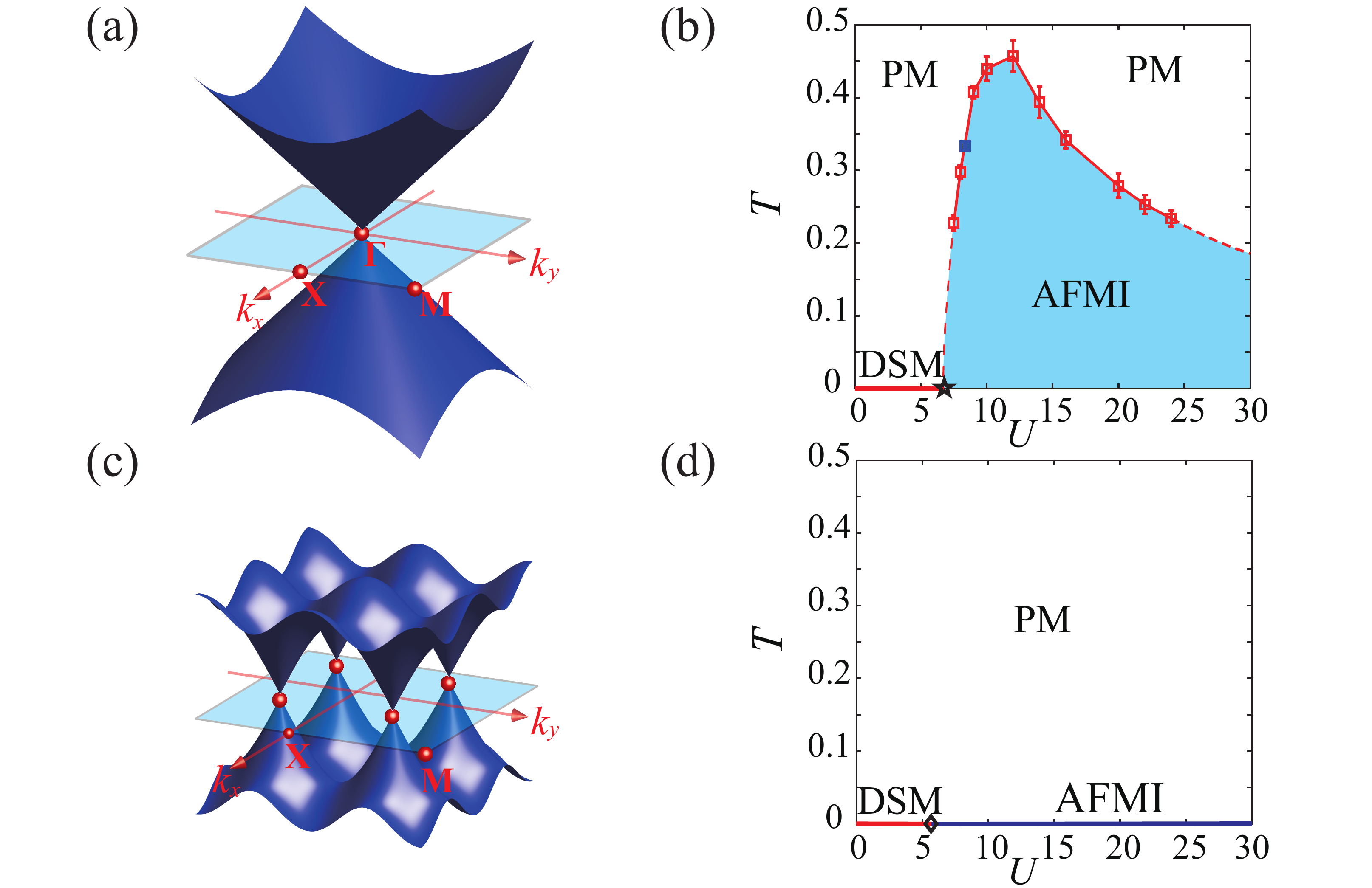}
\caption{The dispersion of (a) SLAC fermion and (c) free $\pi$-flux model in the first BZ. 
The $U$-$T$ phase diagram of (b) SLAC fermion and (d) $\pi$-flux Hubbard model obtained from QMC simulation. In panel (b), the red squares are obtained from the cross of $R$ for different $L$ when scanning $T$ at fix $U=7.5, 8, 9, 10, 12, 14, 16, 20, 22$ and $24$. 
The blue square is obtained from the cross of $R$ for different $L$ when scanning $U$ at fix $T=1/3$. 
The black star denotes the position of QCP in Ref.~\cite{langQuantum2019}.
The red dash line is a guide to the eye. 
In panel (d), the black diamond denotes the position of GN-QCP in Ref.~\cite{otsukaUniversal2016}.
}
\label{fig:fig1}
\end{figure}

The SLAC fermion construction gives rise to a linear energy-momentum relation for all lattice momenta at the non-interacting limit (shown in Fig.~\ref{fig:fig1} (a)), therefore reduces the finite-size effect suffered by other local cousins such as the honeycomb and $\pi$-flux models where only a small region of the Brillouin zone (BZ) displays the relativistic behavior at low-energy. The fundamental difference of the SLAC fermion model compared with its local cousins, i.e., the necessity of avoiding the Nielsen-Ninomiya theorem~\cite{nielsenAbsenceI1981,nielsenAbsenceII1981,nielsenNo-Go1981} by violating locality on finite size lattices and the assumption that the locality of the Dirac operator is recovered in the thermodynamic limit (TDL), has not be investigated. This means, with the long-range interactions in the SLAC fermion models (the bare interaction is on-site but the long-range hopping mediates long-range interaction), whether the GN transition and the symmetry-breaking phases obtained thereafter can be discussed as if they were from a purely local model in the origin sense of GN-QCP~\cite{grossDynamical1974,boyackQuantum2021}, are questionable. 

This is the problem solved in this article. Here we show, by means of large-scale QMC simulations, that the
phase diagram of the SLAC fermion model
is fundamentally different from that of a purely local $\pi$-flux Hubbard model on the square lattice.
In particular, we find the antiferromagnetic insulator (AFMI) phase in the SLAC fermion model exists at finite temperatures, which appears to violate the Mermin-Wagner theorem~\cite{merinAbsence1966,hohenbergExistence1967,halperin2019}.
The AFMI phase emerges from the high-temperature paramagnetic (PM) phase via a finite-temperature continuous phase transition, and this continuous transition line smoothly connects to the previously determined GN-QCP at the ground state~\cite{langQuantum2019}.
Contrary to the picture of the Mermin-Wagner theorem, 
where the low-energy fluctuation of the gapless Goldstone mode destroys the long-range order at any finite temperature, 
we find that the magnetic excitations inside the AFMI are gapped without Goldstone mode, although the state spontaneously breaks continuous $SU(2)$ symmetry.

Our results suggest that the long-range interaction in the SLAC fermion model has altered the low-energy effective theory of the interacting Dirac fermions,  the QCP of SLAC fermion model is fundamentally different from that of the local-interaction ones in this way.
We note that examples of nonlocal interaction stabilizing finite-temperature symmetry-breaking phases and giving rise to gapped Goldstone modes at zero-temperature, have also been seen in 1D Ising and SLAC fermion model~\cite{Cannas1995,wangOn2022} and 2D Heisenberg model~\cite{Fisher1972,diesselGeneralized2022,songDynamical2023}, and in dissipative systems such as 1D Ohmic spin chain~\cite{wernerQuantum2005,weberDissipation2022} and 2D dissipative quantum XY models~\cite{zhuLocal2015,zhuQuantum2016}.

\section{Model and Method}
We consider the spin-1/2 SLAC fermion and the $\pi$-flux Hubbard model on the square lattice at half-filling for comparison.

The SLAC fermion Hubbard model has the Hamiltonian
\begin{equation}
H_\text{SLAC}=-t \sum_{i j \sigma}( A_{i j} c_{i a \sigma}^{\dagger} c_{j b \sigma}+\text { h.c.}) + \frac{U}{2}\sum_{i}\sum_{\lambda=a,b}\left(n_{i  \lambda}-1\right)^2,
\end{equation}
where we set $t=1$ as the energy unit, $c_{i a \sigma}^{\dagger}$ and $c_{i b \sigma}$ are the creation and annihilation operators for an electron at unit cell $i$ on sublattices $a, b$ with spin  $\sigma=\uparrow,\downarrow$; 
$n_{i  \lambda}=\sum_{\sigma} c_{i \lambda \sigma}^{\dagger} c_{i \lambda \sigma}$ denotes the local particle number operator at sublattice $\lambda$ of unit cell $i$;
$A_{ij}= i \frac{(-1)^{x} \pi}{L \sin (x \pi / L)} \delta_{y,0} + \frac{(-1)^{y} \pi}{L \sin (y \pi / L)} \delta_{x,0}$ denotes the electron hopping amplitude with $\mathbf{r}\equiv(x,y)=\mathbf{r}_i -\mathbf{r}_j$ standing for the relative distance between two different unit cells $i$ and $j$ ,  $ x = 1,\cdots, L-1$ with $L$ the linear system size. 
The kinetic term of $H_\text{SLAC}$ is known as the SLAC fermion~\cite{drellStrong1976}, and the corresponding single particle spectrum is $\varepsilon(\mathbf{k})=\pm |\mathbf{k}|$, which results in a single linearly dispersing Dirac cone at momentum $\mathbf{\Gamma}=(0,0)$ point, as shown in Fig.~\ref{fig:fig1} (a).  
We observe that on finite-size lattices, the corresponding Fermi velocity is reduced to $\pm 1$ in the BZ. However, the fermion velocity changes sign at the BZ boundary, resulting in a sigularity. The violation of the locality of SLAC fermion represents itself as singular values at the BZ boundary. 
And previous works~\cite{langQuantum2019,tabatabaeiChiral2022} assume the locality of the Dirac operator is recovered at the TDL.

To make a proper comparison with the local model, we also simulate the $\pi$-flux Hubbard model with the Hamiltonian  
\begin{equation}
H_{\pi\text{-Flux}}=-t \sum_{\langle i j \rangle, \sigma} (B_{i j} c_{i \sigma}^{\dagger} c_{j \sigma}+\text { h.c.}) + \frac{U}{2}\sum_{i}\left(n_{i  }-1\right)^2,
\end{equation}
where hopping amplitudes $B_{i,i+\vec{e}_{x}} = 1$  and  $B_{i,i+\vec{e}_{y}} = (-1)^{i_x}$, the position of site $i$ is given as $\mathbf{r}_i = i_{x}\vec{e}_{x}+i_{y}\vec{e}_{y}$,
such arrangement bestows a $\pi$-flux penetrating each square plaquette (the dispersion is given in Fig.~\ref{fig:fig1} (c)). It is known that the $\pi$-flux model has a chiral Heisenberg GN-QCP at $U_c=5.65(5)$~\cite{changQuantum2012,toldinFermionic2015,otsukaUniversal2016,langQuantum2019}, and the AFMI at $U>U_c$ spontaneously breaking the spin $SU(2)$ symmetry with Goldstone mode located at $\mathbf{M}=(\pi,\pi)$ point (see Fig.~\ref{fig:fig1} (d)).

We employ the projection QMC (PQMC)~\cite{assaadWorld2008} method to study the ground-state and dynamical spin correlation functions and the finite temperature QMC (FTQMC)~\cite{blankenbeclerMonte1981a,hirschTwo-dimensional1985a} method to study the temperature dependence of the physical observables.
These results give rise to a consistent and complementary picture. 
For PQMC method, we can measure a physical observable $\langle O \rangle$ according to
$\langle O\rangle=\lim _{\Theta \rightarrow \infty} \frac{\left\langle\Psi_T\left|e^{-\frac{\Theta}{2} {H}} O e^{-\frac{\Theta}{2} {H}}\right| \Psi_T\right\rangle}{\left\langle\Psi_T\left|e^{-\Theta {H}}\right| \Psi_T\right\rangle}$,
where $\Theta$ is the projection length;  $\left|\Psi_T\right\rangle$ is the trial wave function; and $\left|\Psi_0\right\rangle=\lim _{\Theta \rightarrow \infty} e^{-\frac{\theta}{2} {H}}\left|\Psi_T\right\rangle$ is the ground state wave function.
For FTQMC method, $\langle O\rangle$ can be measured according to 
$\langle O\rangle = \frac{\operatorname{Tr}\left[\mathrm{e}^{-\beta H} O \right]}{\operatorname{Tr}\left[\mathrm{e}^{-\beta H}\right]}$,
where $\beta=1/T$ is the inverse of temperature.
We use discrete $\Theta=M\Delta\tau$ ($\beta=M\Delta\tau$) and perform a Trotter decomposition for PQMC (FTQMC) method, 
and set $\Delta\tau=0.1$ and projection time $\Theta  = 2L+10$ for $H_{\text{SLAC}}$ and $\Theta  = L+10$ for $H_{\pi\text{-Flux}}$ when measuring imaginary-time physical quantities, and, in FTQMC method, we set $\Delta\tau=0.01$ for  measurement.
With the aid of particle-hole symmetry, the PQMC and FTQMC for $H_{\text{SLAC}}$ and $H_{\pi\text{-Flux}}$ models are all sign-problem free~\cite{wuSufficient2005,assaadWorld2008,otsukaUniversal2016,langQuantum2019,panSign2022}.
We have simulated the square lattice system with $N=2L^2$ sites and the linear size $L=5,7,\cdots,19$ for $H_{\text{SLAC}}$, while $N=L^2$ sites and the linear size $L=4,8,\cdots,32$ for $H_{\pi\text{-Flux}}$. 
\begin{figure}[htp!]
	\includegraphics[width=\columnwidth]{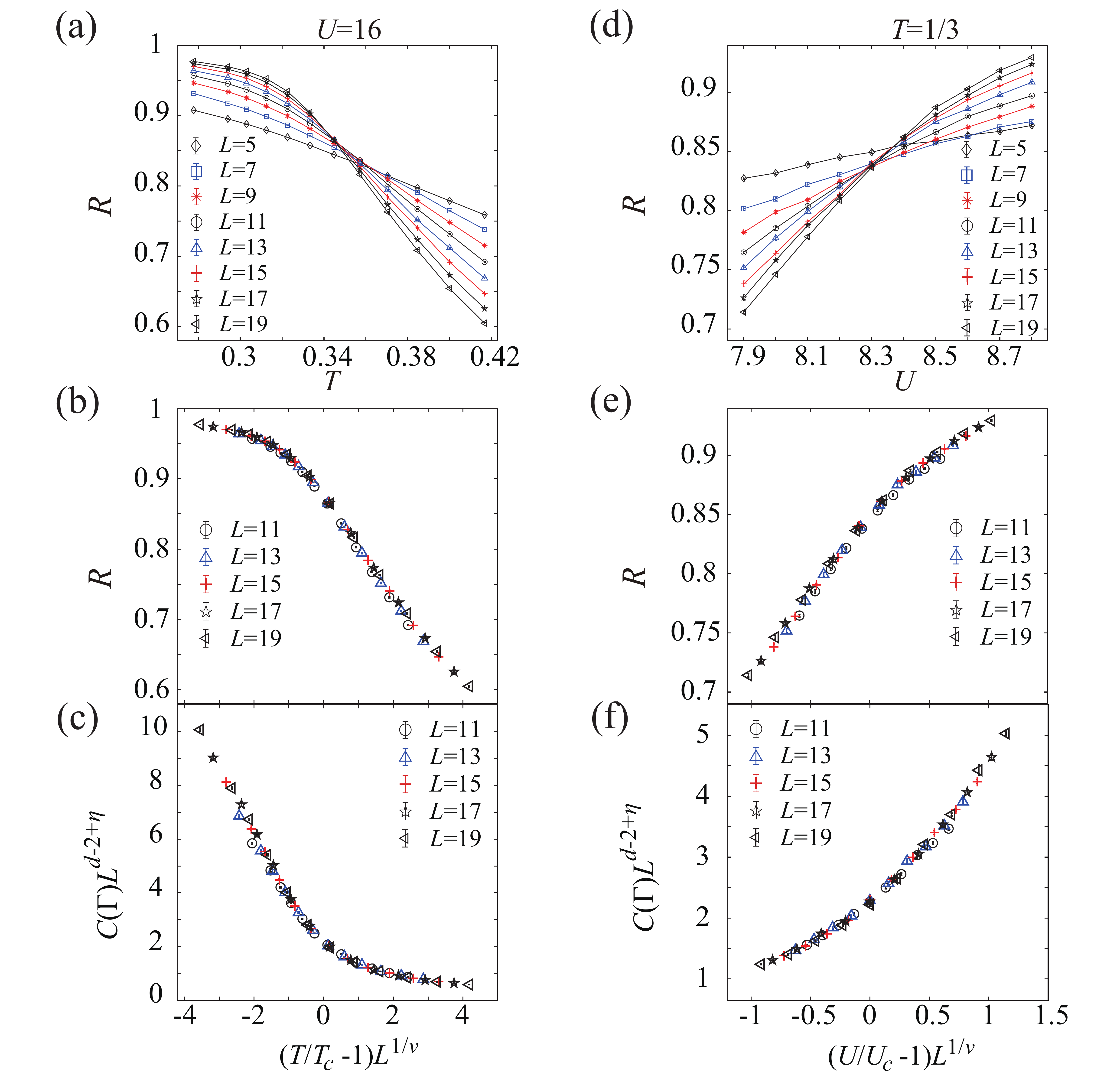}
	\caption{Fixing $U=16$ and scanning $T$, we get (a) the cross and (b) data collapse of correlation ratio $R$, and (c) data collapse of magnetic structure factor $C(\Gamma)$. One can read $T_c=0.34(2)$ from panel (a), extract $\nu=1.0(1)$ from panel (b) and extract $\eta=1.1(1)$ from panel (c).
	Fixing $T=1/3$ and scanning $U$, we get (d) the cross and (e) data collapse of correlation ratio $R$, and (f) data collapse of spin structure factor $C(\Gamma)$. One can read $U_c=8.4(1)$ from panel (d), extract $\nu=1.0(1)$ from panel (e) and extract $\eta=1.2(1)$ from panel (f).}
	\label{fig:fig2}
\end{figure}

\section{Results}
We first reveal the finite temperature continuous phase transition of the AFMI phase in $H_{\text{SLAC}}$, with the phase boundary determined as shown in Fig.~\ref{fig:fig1} (b). Here we use one vertical scan with fixed $U=16$ and varying $T$ and one horizontal scan with fixed $T=1/3$ and varying $U$, to demonstrate the generic behavior.
Fig.~\ref{fig:fig1} (d) are the $U-t$ phase diagram of $\pi$-flux Hubbard model~\cite{otsukaUniversal2016}, we notice that there is no finite temperature phase transition. 

Ref.~\cite{langQuantum2019} investigated the ground state phase diagram of $H_{\text{SLAC}}$. Following their approach, we define the AFMI spin structure factor as
\begin{equation}
C(\mathbf{q})\equiv \frac{1}{L^2} \sum_{ij} e^{i\mathbf{q}\cdot(\mathbf{r}_i-\mathbf{r}_j)} \langle \mathbf{S}_i \cdot  \mathbf{S}_j \rangle,
\end{equation}
where, $\mathbf{S}_i = \frac{1}{2}  c_{ia \sigma}^{\dagger} \boldsymbol{\sigma}_{\sigma \sigma^{\prime}} c_{ia \sigma^{\prime}} - \frac{1}{2}  c_{ib \sigma}^{\dagger} \boldsymbol{\sigma}_{\sigma \sigma^{\prime}} c_{ib \sigma^{\prime}}$ ($\mathbf{S}_i = \frac{1}{2} c_{i \sigma}^{\dagger} \boldsymbol{\sigma}_{\sigma \sigma^{\prime}} c_{i \sigma^{\prime}}$ ) is the fermion spin operator at unite cell (site) $i$ for $H_{\text{SLAC}}$ ($H_{\pi-\text{Flux}}$) and $\boldsymbol{\sigma}$ denotes the Pauli matrices of $SU(2)$ spin. 
For $H_{\text{SLAC}}$ ($H_{\pi-\text{Flux}}$), the AFMI ordering wave vector is $\mathbf{q}=\mathbf{\Gamma}$ ($\mathbf{q}=\mathbf{M}$). 
To locate the thermal phase transition point of $H_{\text{SLAC}}$, we define the renormalization-group invariant correlation ratio $R=1-C(\mathbf{q}+\mathbf{b}_1/L+\mathbf{b}_2/L)/C{(\mathbf{q})}$, 
where $\mathbf{b}_{1,2}$ are the reciprocal lattice vectors~\cite{pujariInteraction2016}.

Fig.~\ref{fig:fig2} (a) and (d) are the correlation ratio $R$ for the two scans as a function of $T$ and $U$, respectively. It is clear that different system sizes have a crossing point both on the $T$ and $U$ axes. With the $T_c=0.34(2)$ at $U=16$ and $U_c=8.4(1)$ at $T=1/3$ obtained, we can further rescale their $x$ axes as $(T/T_c-1)L^{1/\nu}$ and $(U/U_c-1)L^{1/\nu}$ to have good data collapses as shown in Fig.~\ref{fig:fig2} (b) and (e). The collapse successfully give rise to the correlation length exponent $\nu=1.0(1)$ for data in (b) and (e). With the obtained $T_c$, $U_c$ and $\nu$, we can further collapse the AFMI spin structure factor $C(\Gamma)$ near $T_c$, with $C(\Gamma)L^{d-2+\eta}$ and $d=2$. The results are shown in Fig.~\ref{fig:fig2} (c) and (f), and from here, we can further read the anomalous dimension exponent $\eta=1.1(1)$ in the $T$-scan and the $1.2(1)$ in the $U$-scan, of the finite temperature continuous phase transition between the paramagnetic state to AFMI state. In fact, the phase boundary in the Fig.~\ref{fig:fig1} (b), is obtained in this way. We note, the obtained $\eta$ and $\nu$ are indeed consistent with the RG results of 2D Heisenberg model with $1/r^{2.9(1)}$ long-range interaction~\cite{Fisher1972}.

\begin{figure}[htp!]
\includegraphics[width=\columnwidth]{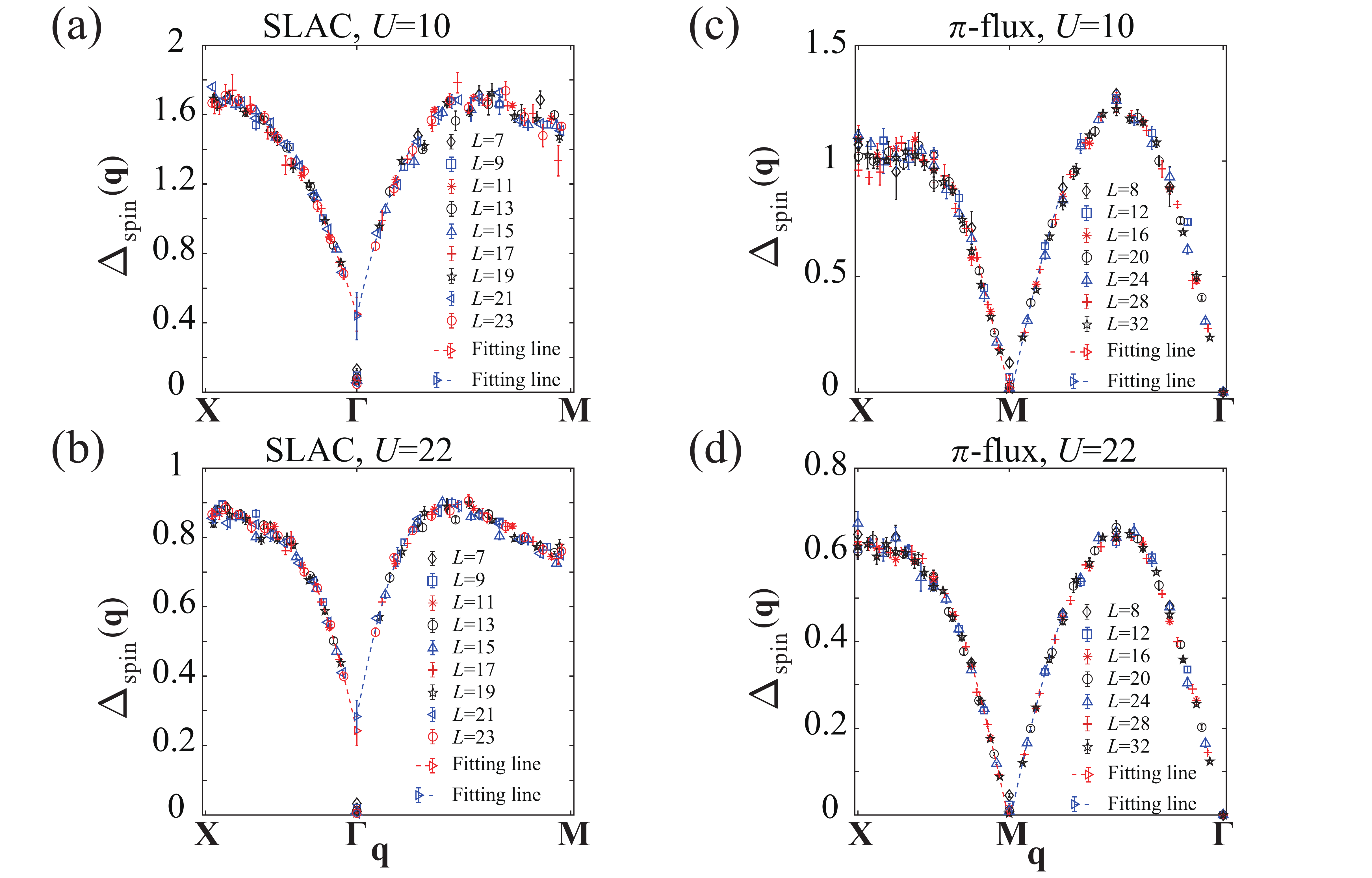}
\caption{The spin excitation gap $\Delta_{\text{spin}}(\mathbf{q})$ along the high-symmetry path in the BZ for (a) $H_{\text{SLAC}}$ with $U=10$, (b) $H_{\text{SLAC}}$ with $U=22$, (c) $H_{\pi-\text{Flux}}$ with $U=10$ and (d) $H_{\pi-\text{Flux}}$ with $U=22$.
The dash lines are obtained by the extrapolation with square polynomial fitting of $\Delta_{\text{spin}}(\mathbf{q})$ at the adjacent momenta to the corresponding AFM wave vectors for different system sizes, and the details of extrapolation could be found in Appendix~\ref{App:B}.
For SLAC, the spin gap $\Delta_\text{spin}=0.45t$ for $U=10$, and  $\Delta_\text{spin}=0.26t$ for $U=22$; while for $\pi$-flux, $\Delta_\text{spin}=0$ for both $U=10$ and $U=22$.
It's clear that there are no gapless Goldstone modes in (a) and (b) and there are gapless spin wave spectra in (c) and (d).}
	\label{fig:fig3}
\end{figure}

The AFMI in the $H_{\text{SLAC}}$ breaks the spin $SU(2)$ symmetry at finite temperature, this is clearly against the requirement of the Mermin-Wagner theorem, which prohibits such transition for 2D systems. The reason of such violation is the violation of the locality in $H_{\text{SLAC}}$ in the first place.
As mentioned, long-range interaction is responsible for such behavior~\cite{diesselGeneralized2022,songDynamical2023,Cannas1995,Fisher1972,weberDissipation2022,zhuLocal2015,zhuQuantum2016}. With the violation of the locality, many of the assumed properties in the symmetry-breaking phase, as well as that of the assumed GN-QCP, have to be reconsidered.
In particular, as we now turn to the dynamic properties of the AFMI and make comparison between the SLAC fermion and the $\pi$-flux models, we find the AFMI in the $H_{\text{SLAC}}$ has no gapless Goldstone modes, whereas the same phase in $H_{\pi-\text{Flux}}$ has them.
This partially explains the apparent violation of Mermin-Wagner theorem, because the theorem asserts that the infrared divergence in the low-energy fluctuation of gapless Goldstone mode destroys the long-range order at any finite temperature.

The results are shown in Fig.~\ref{fig:fig3}, where we have extracted the spin excitation gap $\Delta_{\text{spin}}$ from the dynamic spin correlation functions obtained in PQMC, $C(\mathbf{q},\tau) \sim \exp(-\Delta_{\text{spin}}(\mathbf{q})\tau)$, via fitting the finite size data in their imaginary time decay. The raw data of dynamic spin correlation functions and the fitting procedure are shown in the Appendix~\ref{App:B}. We note since on finite lattice simulation spin-spin correlation at the AFM wavevector is a conserved quantity, resulting in  $\Delta_{\text{spin}}(\mathbf{\Gamma})=0$ for SLAC and $\Delta_{\text{spin}}(\mathbf{M})=0$, therefore one shall look for the asymptotical behavior of $\Delta_{\text{spin}}(\mathbf{q})$ as $\mathbf{q}$ approaches $\mathbf{\Gamma}$ or $\mathbf{M}$~\cite{songDynamical2023,diesselGeneralized2022}.

Fig.~\ref{fig:fig3} (a) and (c) compare the obtained spin gap along the high-symmetry-path of the BZ for $H_{\text{SLAC}}$ and $H_{\pi-\text{Flux}}$ at $U=10$. It is clear that as the system size increases, the $\Delta_{\text{spin}}(\mathbf{q})$ outlines the converged spin wave dispersion for both AFMIs. In the $H_{\text{SLAC}}$ case, the AFM wavevector is at $\mathbf{\Gamma}$, in the vicinity of $\mathbf{\Gamma}$, there is a clear finite energy gap at the scale of 0.4 from the extrapolation of $\mathbf{X} \to \mathbf{\Gamma}$ and $\mathbf{M} \to \mathbf{\Gamma}$, such a large energy gap is clearly not a finite size effect which usually goes as $1/L$ as one is approaching the AFM wavevector with increasing $L$ and it is in sharp contrast with the data in Fig.~\ref{fig:fig3} (c), where in the vicinity of the AFM ordered wavevector at $\mathbf{M}$, the gap is vanishing  (scales as $1/L$ from $\mathbf{X} \to \mathbf{M}$ and $\mathbf{\Gamma} \to \mathbf{M}$) and a gapless Goldstone mode with linear dispersion originated from $\mathbf{M}$ is clearly seen. When we further increase the $U$ to $U=22$ for both models, i.e., deep inside the AFMI phase, the same contrast still present, as shown in Fig.~\ref{fig:fig3} (b) and (d). Therefore, besides the apparent violation of the Mermin-Wagner theorem, the AFMI phase in the $H_{\text{SLAC}}$ also has no Goldstone mode to meet the requirement of spontaneous continuous symmetry breaking. And it means the low-energy effective theory of the AFMI is different between $H_{\text{SLAC}}$ and $H_{\pi-\text{Flux}}$.

\begin{figure}[htp!]
\includegraphics[width=\columnwidth]{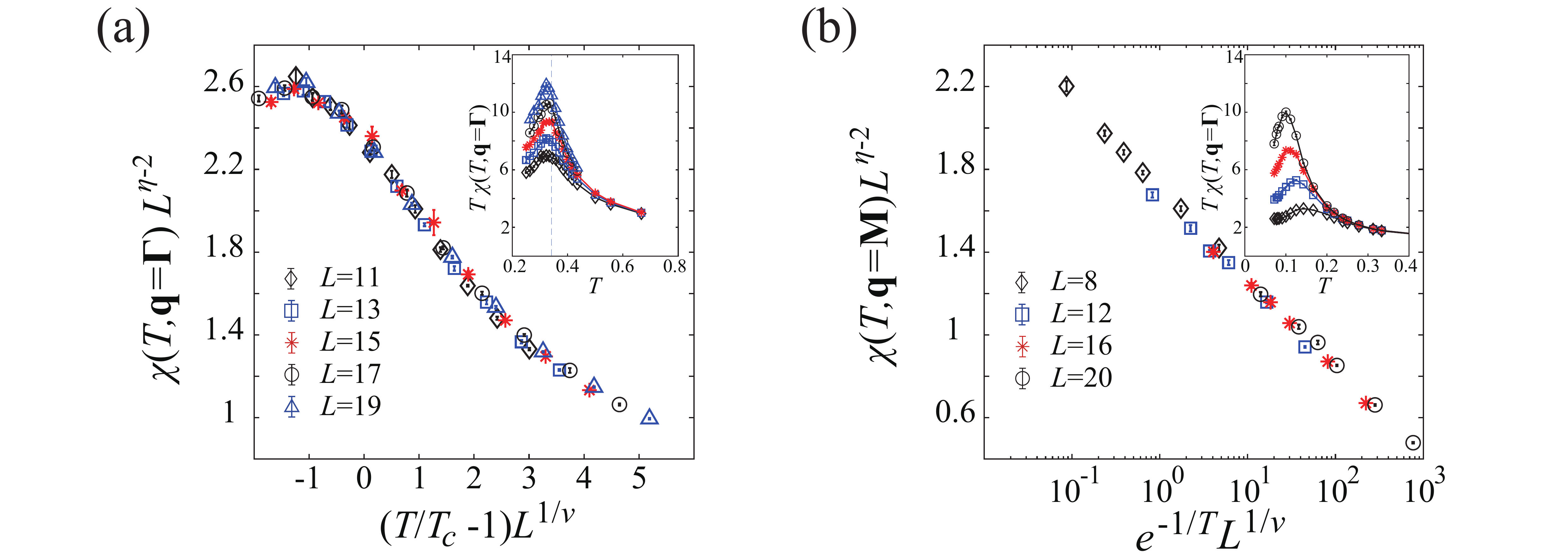}
\caption{The data collapse of magnetic susceptibilities $\chi(T,\mathbf{q})$ for $H_{\text{SLAC}}$ (panel (a)) and $H_{\pi-\text{Flux}}$ (panel (b)) at fixed $U=16$. 
In panel (a), we could collapse magnetic susceptibilities for different system sizes into a smooth curve in terms of the critical temperature $T_c=0.34$ and critical exponents $\nu=1.0$ and $\eta=1.1$ that we extracted from Fig.~\ref{fig:fig2}. 
Inset: The $T\chi(T,\mathbf{\Gamma})$ as function of $T$ shows a peak near $T_c=0.34$.  
In panel (b), we could obtain perfect collapse results according to the renormalized-classical scaling behavior of the 2D O(3) model, where $\eta=0$ and $\nu$ is a nonuniversal exponent.
Inset: The $T\chi(T,\mathbf{M})$ as function of $T$ shows that the peaks at finite size drift towards $T=0$.}
	\label{fig:fig4}
\end{figure}

Besides the dynamic properties, the difference of $H_{\text{SLAC}}$ and $H_{\pi-\text{Flux}}$ also manifests in their thermodynamic response functions, to this end, we compute their magnetic susceptibilities $\chi$ as a function of $T$,
\begin{equation}
	\chi(T,\mathbf{q})=\frac{1}{T}  \left[\frac{1}{L^2}  \sum_{i,j} e^{i\mathbf{q}\cdot(\mathbf{r_i}-\mathbf{r_j})} \langle \mathbf{S}_i \cdot \mathbf{S}_j \rangle- \sum_{i} \langle \mathbf{S}_i \cdot \mathbf{S}_{i'} \rangle \right],
\label{eq:eq6}
\end{equation}
where $i'$ denotes the site which is farthest from site $i$ in real space, and again we chose $\mathbf{q}=\mathbf{\Gamma}$  ($\mathbf{q}=\mathbf{M}$) for $H_{\text{SLAC}}$ ($H_{\pi-\text{Flux}}$), respectively.
Here, we use the equal-time susceptibility instead of the zero-frequency susceptibility, because the former is easier to compute (it has a smaller statistical error due to a smaller variance in FTQMC simulations).
The two susceptibilities exhibits the same scaling behavior near the finite-temperature critical point, because the imaginary fluctuation is irrelevant at the classical critical point.
Moreover, we plot $T\chi(T,\mathbf q)$ instead of $\chi(T,\mathbf q)$ because the former converges to a finite value in the zero-temperature limit.
The $T\chi(T,\mathbf{q})$ data are shown in the insets of Fig.~\ref{fig:fig4}, where the two panels are for $H_{\text{SLAC}}$ and $H_{\pi-\text{Flux}}$ at $U=16$, respectively.
There is clearly a peak in the inset of panel (a), whose location converges to $T_c=0.34(2)$ and amplitude diverges in the thermodynamical limit; whereas in the case of $H_{\pi-\text{Flux}}$ in panel (b), the peak drifts towards $T\to 0$ as the system size increases.
Indeed, in panel (a), it is shown that the susceptibility in the SLAC model satisfies the universal scaling form $\chi(T,\mathbf q)L^{\eta-2}=f(tL^{1/\nu})$, where $f$ is a universal function, and $t=T/T_c-1$ is the reduced temperature.
Similarly, in panel (b), it is shown that the susceptibility in the $\pi$-flux model also satisfies the scaling form $\chi(T,\mathbf q)L^{\eta-2}=f(tL^{1/\nu})$, where the reduced temperature is defined as $t=e^{-1/T}$ for this zero-temperature critical point, and $\nu$ is a nonuniversal exponent.
Such a scaling form can be deduced from the renormalized-classical scaling behavior of the 2D O(3) model~\cite{Chakravarty1988} and $\eta=0$ in this case.
The peaks and scaling behaviors are consistent with a finite-temperature and a zero-temperature phase transition, respectively.
This again means that the continuous spin $SU(2)$ symmetry is broken at finite temperature for the $H_{\text{SLAC}}$ and at zero temperature of the $H_{\pi-\text{Flux}}$.

\section{Discussion}
It is well-known that the violation of locality in quantum many-body systems, either in the form of spatial long-range interaction~\cite{Fisher1972,Cannas1995,songDynamical2023,diesselGeneralized2022,wangOn2022} or the dissipative interaction that introduces the long-range retarded interaction in temporal direction~\cite{wernerQuantum2005,weberDissipation2022,zhuLocal2015,zhuQuantum2016}, will fundamentally change the universalities of the original short-range models and give rise to different behaviors, for example, diverging dynamic exponent~\cite{zhuLocal2015,zhuQuantum2016} and finite temperature order with spontaneous continuous symmetry breaking~\cite{wernerQuantum2005,weberDissipation2022}. Such general expectation, however, has not been explicitly shown in the fermionic systems at 2D, due primarily to the associated computational and analytic complexities.

Here we take a different angle of the active research on the SLAC fermion Hubbard model and the assumed GN-QCPs~\cite{langQuantum2019,tabatabaeiChiral2022}. We find although such an intelligent lattice construction indeed give rise to the linear energy-momentum relation for all lattice momenta at the non-interacting limit -- therefore greatly reduced the notorious finite size effect in QMC simulations, it also introduce unexpected consequences, in that, the interaction-driven AFMI phase in this case is fundamentally different from that on a purely short-range $\pi$-flux Hubbard model on the square lattice. It not only acquires a finite temperature continuous phase transition, which appears to violate the Mermin-Wagner theorem, and the finite temperature critical line smoothly connects to the previously determined GN-QCP, but also exhibits gapped magnetic excitations without gapless Goldstone mode and different thermodynamic responses compared with AFMI in $\pi$-flux Hubbard model. We believe these are the first set of data that explicitly demonstrate the consequence of the violation of the locality in the correlated Dirac fermion systems in 2D, and the low energy effective theory of AFMI and the QCP in SLAC fermion model are different from those of GN-QCP with short-range interactions. 

\section*{Acknowledgements }
We thank Chandra M. Varma and Fakher F. Assaad for valuable discussions on the dissipative quantum many-body models over the years, and thank Zheng Yan for helpful discussion.
Y.D.L. acknowledges support from National Natural Science Foundation of China (Grant No. 12247114).
Y.Q. acknowledges support from the National Natural Science Foundation of China (Grant Nos. 11874115 and 12174068).
Z.Y.M. acknowledges the support from the Research Grants Council of Hong Kong SAR of China (Grant Nos. 17303019, 17301420, 17301721, AoE/P-701/20 and 17309822), the K. C. Wong Education Foundation (Grant No. GJTD-2020-01) and the Seed Funding “Quantum-Inspired explainable-AI” at the HKU-TCL Joint Research Centre for Artificial Intelligence.
X.Y.X. is sponsored by the National Key R\&D Program of China (Grant No. 2021YFA1401400), the National Natural Science Foundation of China (Grants No. 12274289), Shanghai Pujiang Program under Grant No. 21PJ1407200, Yangyang Development Fund, and startup funds from SJTU.
The authors also acknowledge Beijng PARATERA Tech Co.,Ltd.~\cite{paratera} for providing HPC resources that have contributed to the research results reported in this paper.

\appendix

\section{The comparison of correlation ratio between SLAC fermion and $\pi$-flux Hubbard model}
\begin{figure}[htp!]
\includegraphics[width=\columnwidth]{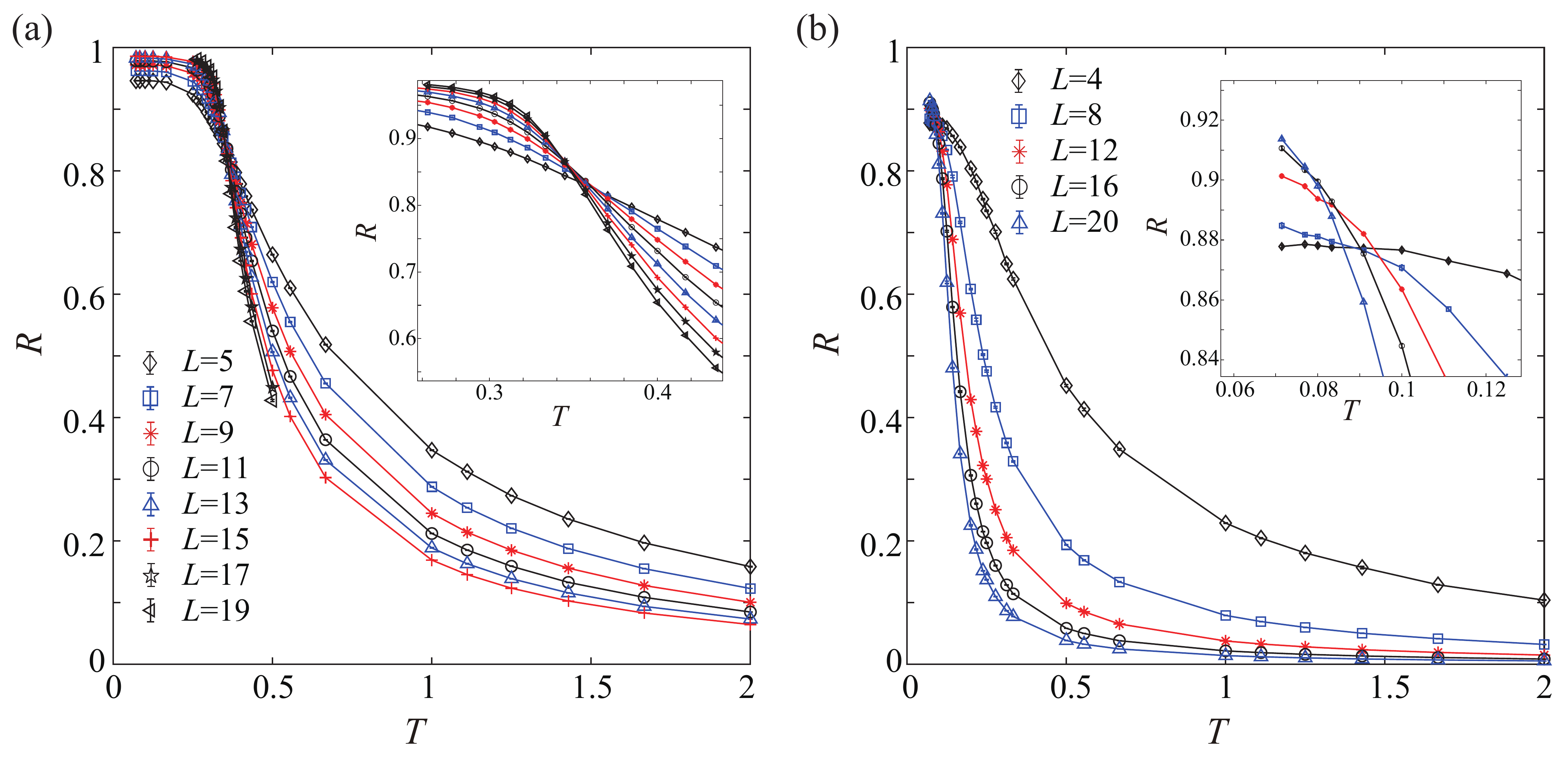}
\caption{The correlation ratio at $U=16$ for $H_{\text{SLAC}}$ (a) and $H_{\pi-\text{Flux}}$ (b). The finite temperature transtion in the former is shown as the crossing point at $T_c=0.34(2)$ (inset of (a)), whereas the zero temperature AFMI phase in the latter is shown as the drift of the finite size crossing points (inset of (b)).}
\label{fig:figS1}
\end{figure}

Here, we compare the SLAC fermion and $\pi$-flux Hubbard model with the correlation ratios obtianed from the spin-spin correlation functions. The parameter is $U=16$ as in Fig.2 (a), (b) and (c). The results are shown in Fig.~\ref{fig:figS1}. We note that there is a clearly converged cross point for different $L$-s in the $H_{\text{SLAC}}$ case with $T_c=0.34(2)$ (inset of (a)), while, in the $H_{\pi-\text{Flux}}$ case, the cross point is not only at smaller $T$ for finite sizes but also drifts towards $T\rightarrow 0$ as $L$ increases (inset of (b)), which is consistent with the fact that the AFMI in  $H_{\pi-\text{Flux}}$ happens at $T=0$.

\section{Dynamic spin correlation function and spin excitation gap }\label{App:B}
\begin{figure}[htp!]
\includegraphics[width=\columnwidth]{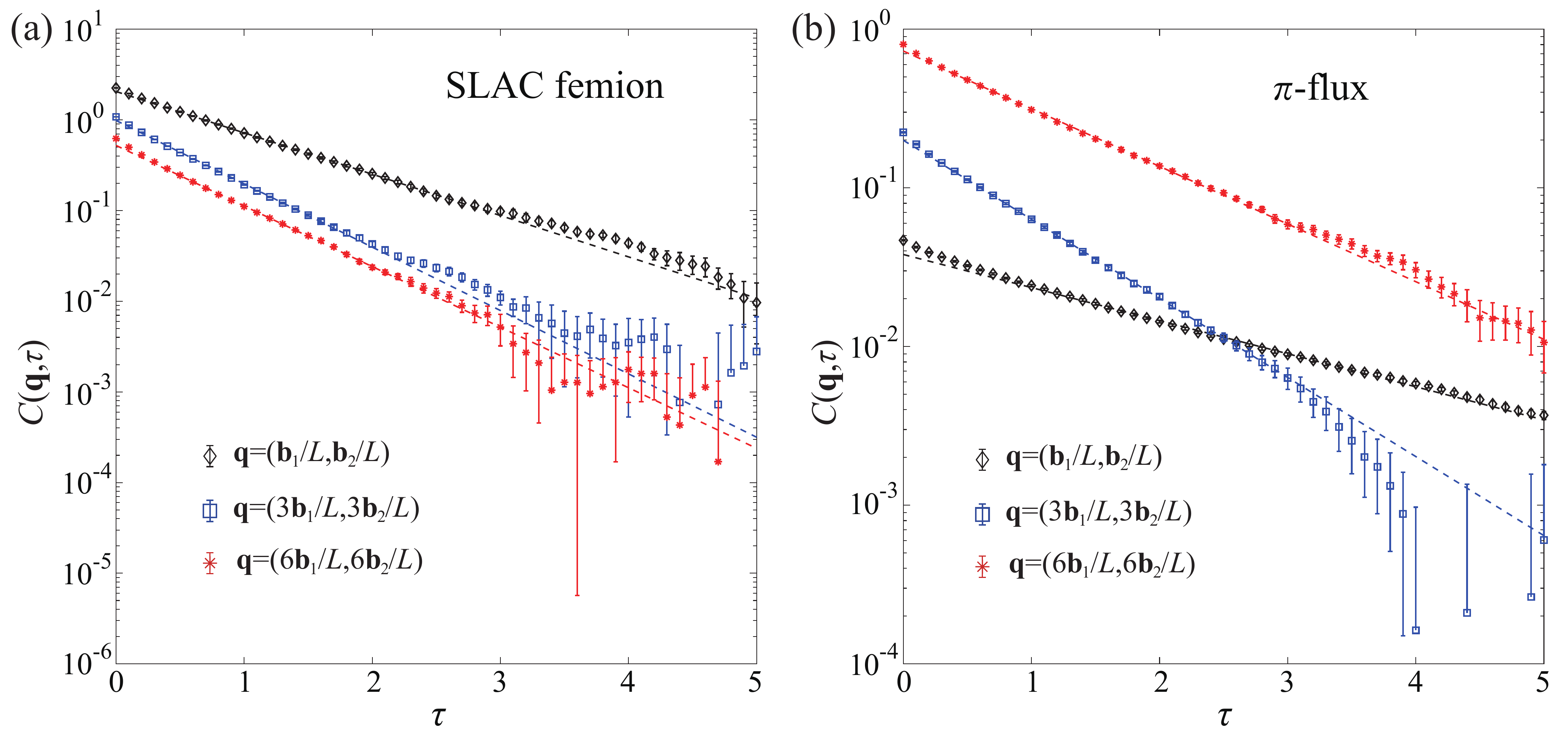}
\caption{Log-plot of dynamic spin correlation function at momenta $\mathbf{q}=(\mathbf{b}_1/L,\mathbf{b}_2/L)$, $\mathbf{q}=(3\mathbf{b}_1/L,3\mathbf{b}_2/L)$ and $\mathbf{q}=(6\mathbf{b}_1/L,6\mathbf{b}_2/L)$ of $L=15$, $U=10$ system for (a) SLAC fermion Hubbard model and (b) $\pi$-flux Hubbard model. The dashed lines are the linear fitting curves from which the $\Delta_{\text{spin}}(\mathbf{q})$ are extracted.}
	\label{fig:figS2}
\end{figure}

To obtain the spin excitation gap $\Delta_\text{spin}(\mathbf{q})$, as shown in the Fig. 3 of the main text, we first calculate the dynamic spin correlation functions $C(\mathbf{q},\tau)\equiv \frac{1}{L^2} \sum_{ij} e^{i \mathbf{q} \cdot\left(\mathbf{r}_i-\mathbf{r}_j\right)}\left\langle\mathbf{S}_{i}(\tau) \cdot \mathbf{S}_{j}(0)\right\rangle$, and then extract the spin excitation gap as $C(\mathbf{q}, \tau) \sim \exp \left(-\Delta_{\operatorname{spin}}(\mathbf{q}) \tau\right)$. Fig.~\ref{fig:figS2} show the exemplary data at three different momenta for $L=15$ and $U=10$ (the same data set of Fig.3 (a) and (b) in the main text) for the $H_{\text{SLAC}}$ (a) and $H_{\pi-\text{Flux}}$ (b) cases. In both cases, the exponential decay in the imaginary time is clear and the fitting can be carried out readily.

\section{The Trotter error analysis and benchmark with ED}
As mentioned in the main text, the Trotter decomposition in the imaginary time in the QMC introduces a systematic error at the scale of $O((\Delta\tau)^2)$. Here we show that our choices of the $\Delta\tau$ are small enough such that for the finite size systems we can access, the convergence of the physical observables are already obtained.

\begin{figure}[htp!]
\includegraphics[width=\columnwidth]{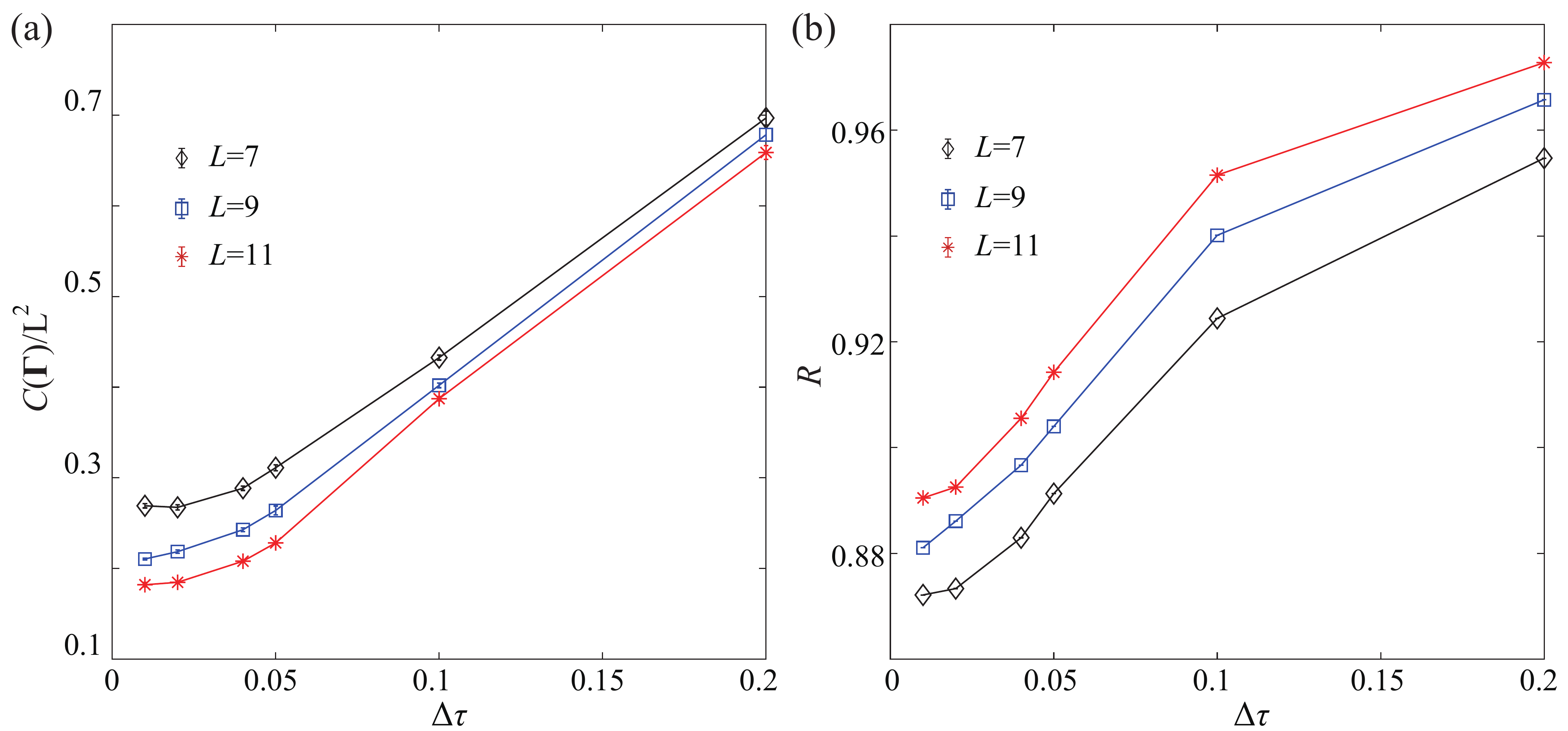}
\caption{(a) $C(\mathbf{\Gamma})/L^2$ and (b) $R$ with respect to the time slice interval $\Delta\tau$. These results are obtained at $T=0.33$ near $T_c$. }
	\label{fig:figS3}
\end{figure}

The behavior of the Trotter error as function of $\Delta\tau$ is shown in Fig.~\ref{fig:figS3} for $H_{\text{SLAC}}$  with FTQMC method. 
We show two physical observables, the square of magnetization $C(\mathbf{\Gamma})/L^2$ and correlation ratio $R$, their definitions are given in the main text.
We choose parameters $U=16$ and $T=0.33$ near $T_c$, and notice that $C(\mathbf{\Gamma})/L^2$ and $R$ are all converged at $\Delta\tau=0.01$ which is the imaginary time discretization we used.
For PQMC method used for calculating the dynamic spin correlation function, we set $\Theta=2L+10$ and $\Delta\tau=0.1$, which is the same discretization used in the Ref.~\cite{langQuantum2019}, as shown there, this value is sufficient to achieve convergent and error controllable numerical results for$H_{\text{SLAC}}$. 

\begin{figure}[htp!]
\includegraphics[width=\columnwidth]{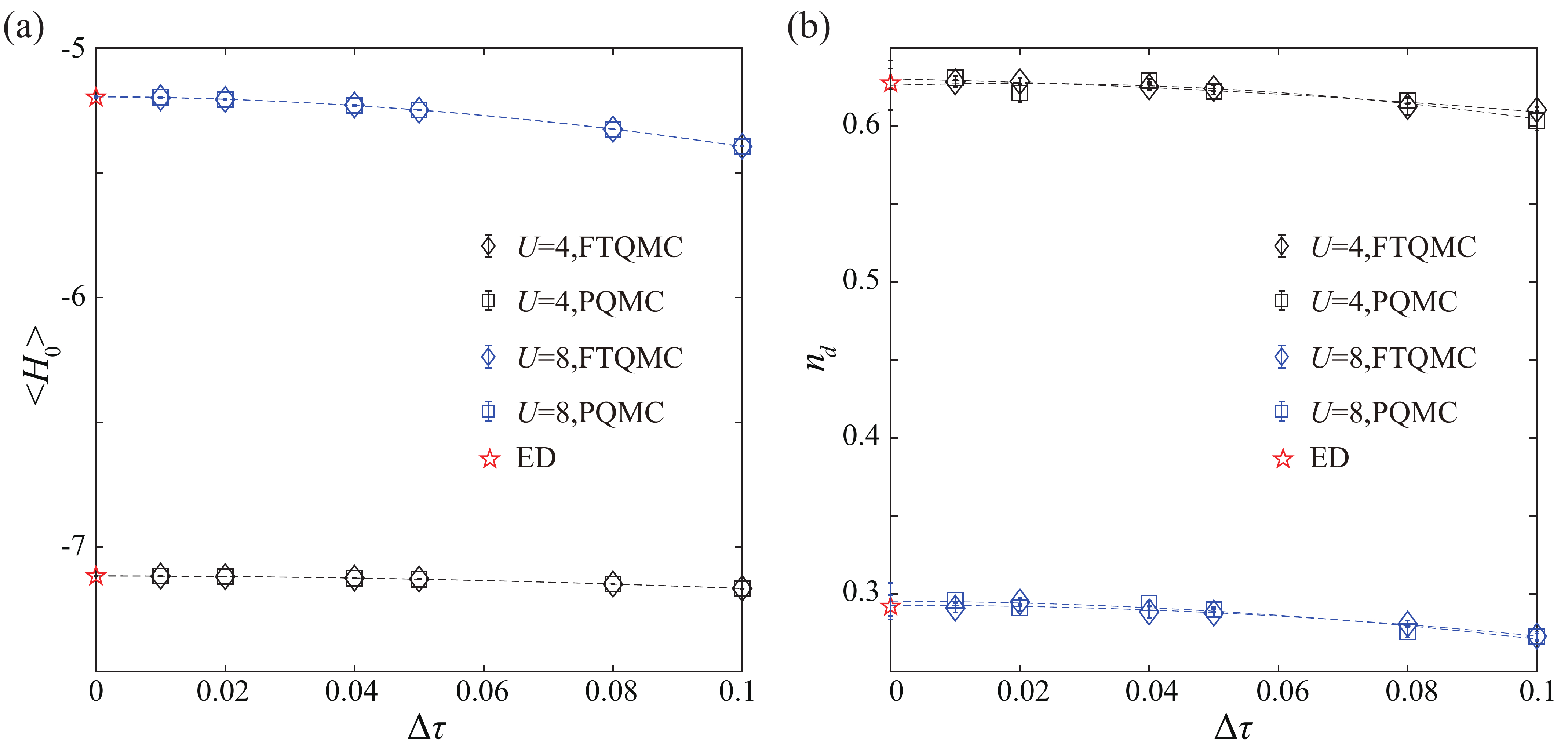}
\caption{ $\Delta\tau$ extrapolation of (a) $\langle H_0 \rangle$ and (b) $n_d$. For PQMC, we set projection time $\Theta  = 100$, while for FTQMC, we set the inverse of temperature $\beta =300$. The dash lines are obtained by square polynomial fitting through the corresponding data.}
	\label{fig:figS4}
\end{figure}

In Fig.~\ref{fig:figS4}, we further show the benchmark of kinetic energy $\langle H_0 \rangle = \langle -t \sum_{i j \sigma}( A_{i j} c_{i a \sigma}^{\dagger} c_{j b \sigma}+\text { h.c.}) \rangle $ and double occupancy $n_d=\sum_{i\lambda}\langle n_{i
\lambda} \rangle$ between exact diagonalization (ED) results and two different QMC estimates for a 6-site ($L_x=3$, $L_y=1$) $H_{\text{SLAC}}$ system at $U=4$ and $8$. 
The extrapolations of the data to $\Delta\tau\rightarrow 0$ are consistent with the ED results within error bars.

\section{The $1/L$ extrapolation of spin excitation gap}
\begin{figure}[htp!]
\includegraphics[width=\columnwidth]{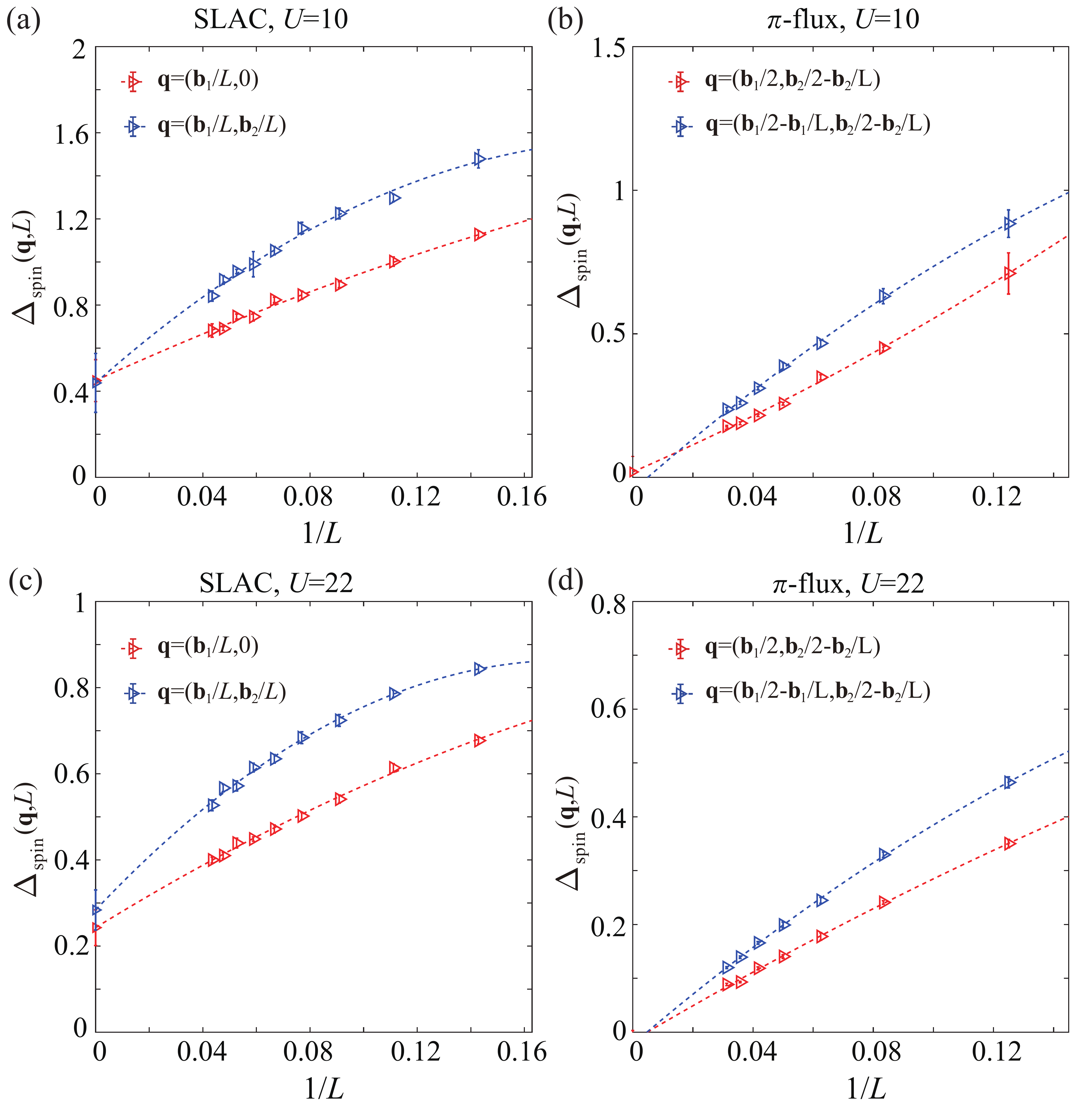}
\caption{ $\Delta_\text{spin}(\mathbf{q},L)$ extrapolation of $H_{\text{SLAC}}$ at (a) $U=10$ and (c) $U=22$, and of $H_{\pi-\text{Flux}}$ at (b) $U=10$ and (d) $U=22$. The dash lines are the quadratic fitting.}
	\label{fig:figS5}
\end{figure}

In the Fig.3 of the main text, we plot the extrapolating lines to claim there is no Goldstone mode in $H_{\text{SLAC}}$, while there indeed is in $H_{\pi-\text{Flux}}$.
Here, we show the details of extrapolations.
As shown in Fig.~\ref{fig:figS5}, we extrapolate the spin excitation gap $\Delta_\text{spin}(\mathbf{q},L)$ to thermodynamic limit.
For $H_{\text{SLAC}}$, as shown in Fig.~\ref{fig:figS5} (a) and (c), we extrapolate $\Delta_\text{spin}(\mathbf{q},L)$ along momenta $\mathbf{q}=\left(\mathbf{b}_1 / L, 0\right)$ (the path of $\mathbf{X}\to \mathbf{\Gamma}$) and $\left(\mathbf{b}_1 / L, \mathbf{b}_2 / L\right)$ (the path of $\mathbf{M} \to \mathbf{\Gamma}$) with quadratic function in $1/L$ for $U=10$ and $22$.
The results clearly show a finite gap at $\mathbf{\Gamma}$ point.
While, the same analysis for $H_{\pi-\text{Flux}}$, the gaps go to zero at $\mathbf{M}$ point, as shown in Fig.~\ref{fig:figS5} (b) and (d), along the paths of $\mathbf{X} \to \mathbf{M}$ and $\mathbf{\Gamma} \to \mathbf{M}$, respectively.

\bibliographystyle{apsrev4-2}
\bibliography{main-slac}

\end{document}